\documentclass[reprint,superscriptaddress,amsmath,amssymb,aps,prb]{revtex4-1}

\usepackage{graphicx}
\usepackage{dcolumn}
\usepackage{bm}
\usepackage{braket}
\usepackage{mathtools}
\usepackage{dsfont}
\usepackage{url}
\usepackage[hidelinks = true]{hyperref}
\usepackage{amsmath}
\usepackage{booktabs}
\usepackage{xcolor}

\begin{document}

\preprint{APS/123-QED}	
	\title{Quantum thermal transistor in superconducting circuits}
	\author{Marco Majland}
	\affiliation{Department of Physics and Astronomy, Aarhus University, Ny Munkegade 120, 8000 Aarhus C, Denmark}
	\author{Kasper Sangild Christensen}
	\affiliation{Department of Physics and Astronomy, Aarhus University, Ny Munkegade 120, 8000 Aarhus C, Denmark}
	\author{Nikolaj Thomas Zinner}
	\email{zinner@aias.au.dk}
	\affiliation{Department of Physics and Astronomy, Aarhus University, Ny Munkegade 120, 8000 Aarhus C, Denmark}
	\affiliation{Aarhus Institute of Advanced Studies, Aarhus University, H{\o}egh-Guldbergs Gade 6B, 8000 Aarhus C, Denmark}
		
	\date{\today}
	
	\begin{abstract}
		Logical devices based on electrical currents are ubiquitous in modern society. However, 
		digital logic does have some drawbacks such as a relatively high power consumption. It is
		therefore of great interest to seek alternative means to build logical circuits that can 
		either work as stand-alone devices or in conjunction with more traditional electronic 
		circuits. One direction that holds great promise is the use of heat currents for 
		logical components. In the present paper, we discuss a recent abstract proposal for 
		a quantum thermal transistor and provide a concrete design of such a device using 
		superconducting circuits. Using a circuit quantum electrodynamics
		Jaynes-Cummings model, we propose a three-terminal device that allows heat transfer
		from source to drain, depending on the temperature of a bath coupled at the gate modulator, 
		and show that it provides similar properties to a conventional semiconductor transistor.
	\end{abstract}

	\maketitle
	
	\setlength\parindent{0pt}
	
	\graphicspath{ {./graphics/} }

	\section{Introduction}
	An inevitable requirement for logical computational devices is the 
	ability to control signals. The most well-known case is that of 
	classical electronic integrated circuits that are based on (semiconductor)
	transistors, a component capable of transmitting or blocking the flow 
	of current based on whether a gate voltage is in the {\it on}- or {\it off}-state.
	While electronics has great advantages, other means of signal transport are 
	under study to extend the versatility and applications of future technologies. 
	A promising path for developing alternative logical devices is to use controlled
	heat currents (phonon transport) to developed thermal components 
	\cite{RevModPhys.78.217,ROBERTS2011648,RevModPhys.84.1045,benenti2017}. 
	Some prominent results of this pursuit are thermal diodes 
	\cite{PhysRevLett.88.094302,PhysRevLett.93.184301,doi:10.1063/1.2191730,chang2006,martinez2015,wang2017}, 
	thermal transistors \cite{PhysRevLett.116.200601,PhysRevE.98.022118}, and related
	designs \cite{PhysRevLett.99.177208,PhysRevLett.101.267203,PhysRevLett.119.090603,PhysRevE.80.061115,PhysRevLett.74.1504}.
	In particular, quantum spin systems coupled to thermal baths have shown promise for the realization 
	of these components 
	\cite{PhysRevB.79.014207,PhysRevB.80.172301,PhysRevE.92.062120,PhysRevE.99.032136}.
	
	A way to realize two-level (spin) systems is to use so-called 
	artificial atoms based on superconducting circuits \cite{2011RPPh...74j4401B,devoret2013,Wendin_2017,GU20171}.
	In the present work, we present a quantum thermal transistor design that is based on 
	superconducting circuits. Our proposal is inspired by the recent work 
	of Guo, Liu, and Yu \cite{PhysRevE.98.022118} in which an abstract 
	model of such a transistor is proposed and discussed. A key ingredient 
	in the model of \cite{PhysRevE.98.022118} is the presence of both 
	two- and three-level systems coupled to heat baths. Since superconducting
	circuits realize non-linear oscillators with several levels, it is possible 
	to realize couplings between qubits and qutrits in such 
	systems \cite{PhysRevA.93.053838,PhysRevLett.104.163601,2018arXiv180204299B,PhysRevApplied.11.014053}. 
	Here we propose a concrete realization of a superconducting circuit that can 
	perform the tasks of a thermal transistor. This requires modifications 
	to the original abstract proposal of \cite{PhysRevE.98.022118} that we will 
	discuss below.
	The transistor is realized as a thermal transistor where the exchanged signal between the two transistor terminals will be in the form of heat. Specifically, the heat will be exchanged between two thermal baths, a source and a drain, through a three-level system (qutrit) and modulated by a third terminal. The third terminal is implemented as a qubit whose population is dictated by a third modulating thermal bath. By controlling the temperature of the modulating bath, one may effectively switch on and off the heat current between the source and drain terminals. The effect of amplification of the heat signal is also investigated. Specifically, the degree to which heat signals may be amplified through the transistor turns out to critically depend on the anharmonicity of the superconducting artificial atoms. Thus, with current state-of-the-art transmons as used in this article, the relatively low anharmonicities weaken the degree of amplification, as will be demonstrated.
	The analysis of the thermal transistor will be as follows. The first step is to understand the dynamics of the coupled qubit/qutrit superconducting circuit (dubbed main circuit). This will be done using elementary circuit analysis. From this point, we will investigate the coupling between the main circuit and the three thermal baths described perturbatively using the theory of open quantum systems within the Lindblad formalism. This allows for the calculations of the heat currents between the source and drain terminals using quantum thermodynamics. 
	Throughout the article, we will use natural units such that $\hbar = c = k = 1$, where $\hbar$ is Planck's constant, $c$ is the speed of light in vacuum and $k$ is Boltzmann's constant, respectively.
	
	\section{Conceptual model and circuit Hamiltonian}
	\label{sec:main_circuit_Hamiltonian}
	\begin{figure}[h]
		\centering
		\includegraphics[width=1.0\columnwidth]{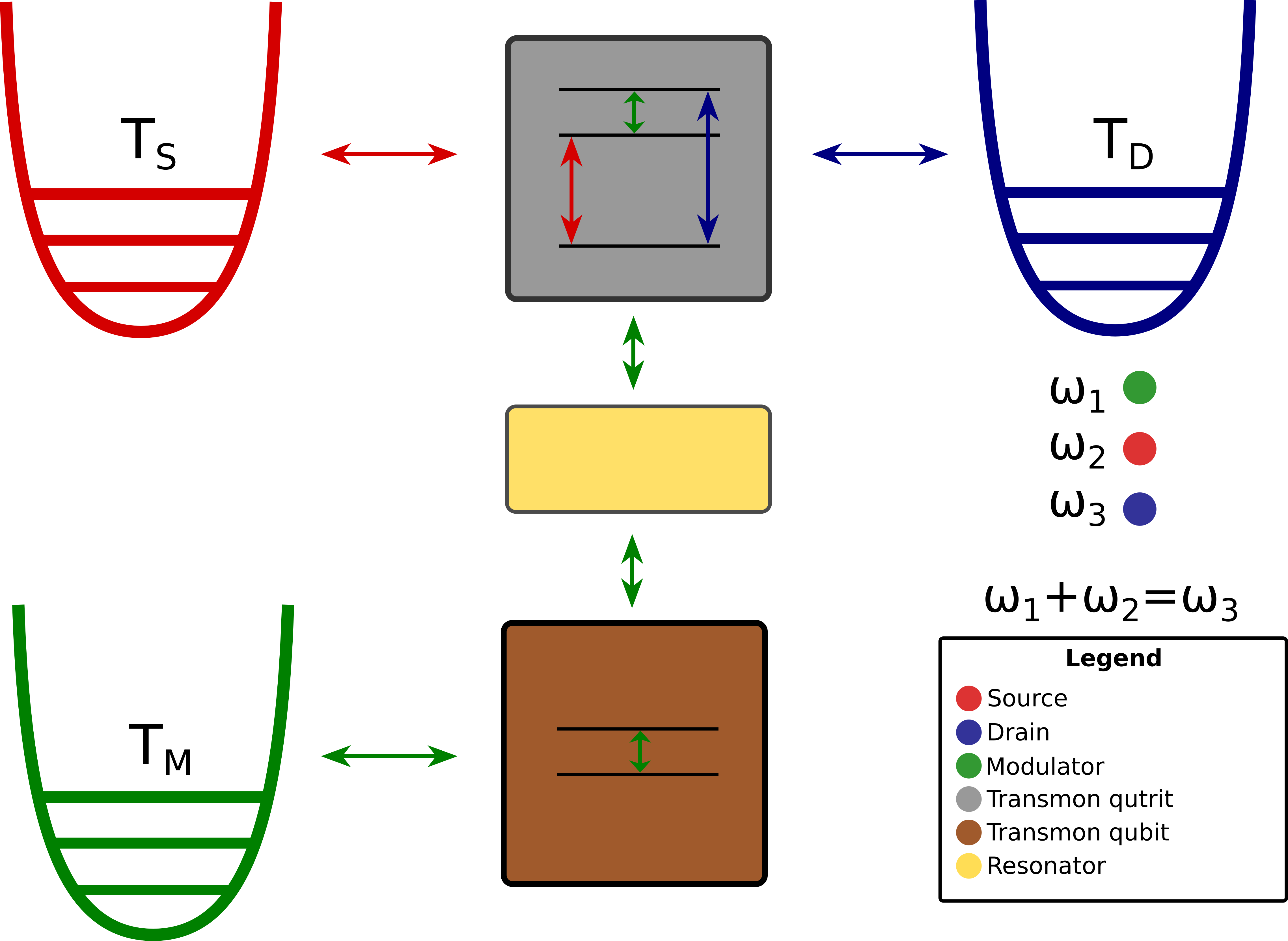}
		\def\svgwidth{\textwidth}
		\caption{Conceptual model of the thermal transistor. Superconducting circuit diagram is presented in Fig. \ref{fig:total_circuit_with_labels}.}
		\label{fig:conceptual_model}
	\end{figure}
\begin{figure}[h]
		\centering
		\includegraphics[width=1.0\columnwidth]{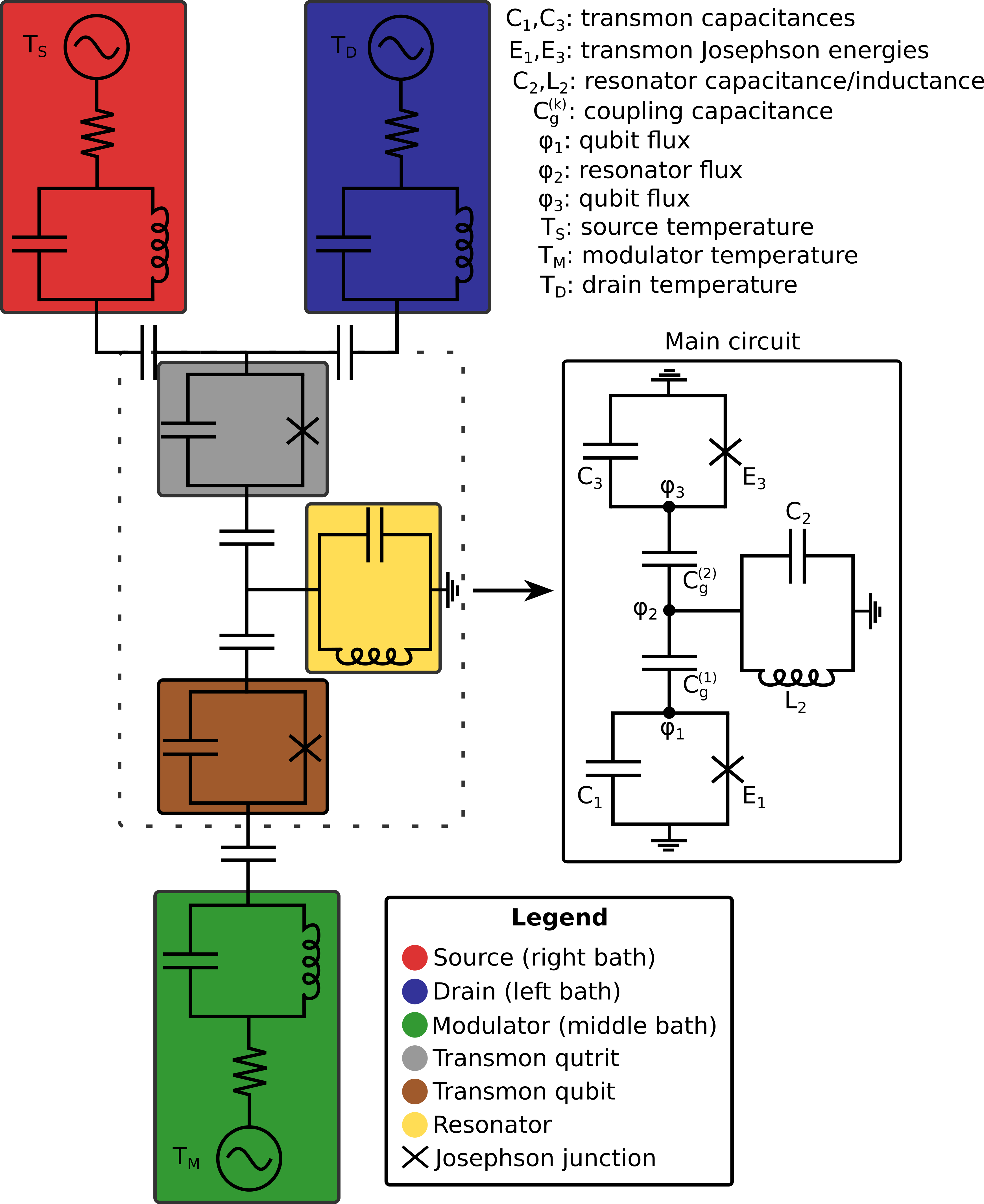}
		\def\svgwidth{\textwidth}
		\caption{Total circuit diagram (left) and a detailed version of the main circuit with parameter descriptions (right). As can be seen, the circuit encapsulated in red is the source terminal, the blue circuit is the drain terminal and the green circuit is the modulating terminal. Heat is exchanged between the source and drain terminal depending on the main circuit dynamics.}
		\label{fig:total_circuit_with_labels}
	\end{figure}

	The circuit implementation of the thermal transistor is depicted in Fig. \ref{fig:total_circuit_with_labels} along with a conceptual model of the circuit in Fig. \ref{fig:conceptual_model}. In Fig \ref{fig:conceptual_model}, the three thermal oscillators at temperatures $T_{S}, T_{D}$ and $T_{M}$ model the thermal baths of the source, drain and modulator terminals. The colors indicate resonant frequencies for both the oscillators and the transmons. The conceptual idea of the transistor is as follows. The exchanged signal between the two terminals (source and drain reservoir/oscillator) is in the form of heat current. This exchange signal will be modulated by a gate terminal (modulator reservoir/oscillator). The source reservoir may interact with the qutrit and excite its first state. The qubit population is dictated by the temperature of the modulating bath and interacts with the qutrit through a resonator. Given a first level excitation of the qutrit, the qubit may further excite the qutrit to its second level, allowing for interaction between the qutrit and the drain reservoir. Thus, one may effectively control the heat flow between the source and drain terminals by modulating the temperature of $T_{M}$. The key interaction in the above mechanism is the interaction between the qubit and the qutrit. In the following, we demonstrate how such an interaction may be engineered in a superconducting circuit architecture.\\
	The thermal transistor circuit may be divided into two subcircuits, the main circuit and the thermal baths, see Fig. \ref{fig:total_circuit_with_labels}.
	In the following, we will analyze the main circuit dynamics and demonstrate its equivalence to the desired dynamics of the abstract thermal transistor Hamiltonian, as presented in \cite{PhysRevE.98.022118}. The main circuit will be composed of two transmons coupled through a resonator through which they may interact through photonic exchange \cite{PhysRevA.76.042319,Wendin_2017}. In their lumped element representations, the transmons are modelled as anharmonic circuits with Josephson junctions and the resonator as an LC circuit \cite{PhysRevA.76.042319}.
	Following the conventional procedure of quantum electromagnetic circuit analysis \cite{2016arXiv161003438V}, we define the generalized flux coordinate 
	$\phi(t) = \int_{-\infty}^{t}V(t')dt'$\footnote{The circuit is assumed to be relaxed as $t\rightarrow -\infty$.}. One may find that $\phi_1, \phi_2$ and $\phi_3$ sufficiently describe the main circuit dynamics as illustrated in Fig. \ref{fig:total_circuit_with_labels}. Each flux describes the degree of freedom for the lower transmon, the resonator and the upper transmon, respectively. Using the detailed circuit analysis presented in appendix \ref{sec:main_circuit_analysis}, we obtain a quantized, step operator Hamiltonian of the main circuit,
	\begin{equation}
		\begin{aligned}
			H = &\alpha_{1}\Big(a_{1}^{\dagger}a_{1} + \frac{1}{2}\Big) - \beta_{1}\Big(a_{1} + a_{1}^{\dagger}\Big)^4 + \alpha_{3}\Big(a_{3}^{\dagger}a_{3} + \frac{1}{2}\Big)\\
			&- \beta_{3}\Big(a_{3} + a_{3}^{\dagger}\Big)^4 + \omega_{r}\Big(b_{r}^{\dagger}b_{r} + \frac{1}{2}\Big)\\
			&- g_{1}\Big(a_{1}^{\dagger}b_{r}^{\dagger} + a_{1}b_{r} - a_{1}^{\dagger}b_{r} - a_{1}b_{r}^{\dagger}\Big)\\
			&- g_{3}\Big(a_{3}^{\dagger}b_{r}^{\dagger} + a_{3}b_{r} - a_{3}^{\dagger}b_{r} - a_{3}b_{r}^{\dagger}\Big).
			\label{eq:non_truncated_circuit_Hamiltonian}
		\end{aligned}
	\end{equation}
	with $\alpha_{1} = \sqrt{8E_{C1}E_{J1}}$, $\beta_{1} = \frac{E_{C1}}{12}$, $\alpha_{3} = \sqrt{8E_{C3}E_{J3}}$, $\beta_3 = \frac{E_{C3}}{12}$, $\omega_r = 4\sqrt{E_{C2}E_{L2}}$, $g_{1} = C_{12}^{-1}\Big(\frac{8E_{J1}E_{L2}}{E_{C1}E_{C2}}\Big)^{1/4}$ and $g_{3} = C_{23}^{-1}\Big(\frac{8E_{J3}E_{L2}}{E_{C2}E_{C3}}\Big)^{1/4}$ with circuit parameters described in Fig. \ref{fig:total_circuit_with_labels}. Operators $a_{i},i\in\{1,3\}$ parametrize the lower and upper transmons, respectively, and $b_{r}$ parametrizes the resonator.
	\subsection{Truncation of the Hamiltonian}
	In order to obtain the desired dynamics, we must truncate the system to the desired degrees of freedom. Specifically, the lower transmon will act as a qubit and the upper transmon as a qutrit. By truncating Eq. \eqref{eq:non_truncated_circuit_Hamiltonian} to these levels, we obtain the Jaynes-Cummings (JC) Hamiltonian. Define $\omega_1$ as the qubit excitation energy, $\omega_2$ as the first excitation energy of the qutrit and $\omega_3$ the second excitation energy.  Since the second excitation level of the qutrit must be dictated by the population of the modulating qubit, we require these two energies to be resonant. For the constraint in question to be fulfilled, we require the resonant condition $\omega_1 + \omega_2 = \omega_3$.\\
	Truncation of the transmon circuit Hilbert spaces to two (qubit) and three (qutrit) dimensions yields a qubit excitation energy of%
	\begin{equation}
		\omega_{1} = \alpha_{1} - 12\beta_{1}.
		\label{eq:qubit_energies}
	\end{equation}
	For the qutrit, one obtains
	\begin{equation}
		\begin{aligned}
			\omega_{2} &= 6\beta_{3} + \sqrt{(\alpha_{3} - 18\beta_{3})^2 + 72\beta_{3}^2}\quad (\textrm{first excited state})\\
			\omega_{3} &= 2\sqrt{(\alpha_{3} - 18\beta_{3})^2 + 72\beta_{3}^2}\quad (\textrm{second excited state}).
			\label{eq:qutrit_energies}
		\end{aligned}
	\end{equation}
	These energy level definitions are depicted in Fig. \ref{fig:conceptual_model}. Eliminating counter-rotating terms (RWA) in Eq. \eqref{eq:non_truncated_circuit_Hamiltonian} yields
	\begin{equation}
		\begin{aligned}
			H = &\omega_{1}\Big(\ket{1}_{1}\bra{1}_{1}\otimes\mathds{1}\Big) + \sum_{i=2}^{3}\omega_{i}\Big(\mathds{1}\otimes\ket{i-1}_{2}\bra{i-1}_{2}\Big)\\
			&+ \omega_rb_{r}^{\dagger}b_{r} + g_{1}\Big(a_{1}^{\dagger}b_{r} + h.c.\Big) + g_{3}\Big(a_{3}^{\dagger}b_{r} + h.c.\Big).
			\label{eq:non_ket_representation_interaction_term}
		\end{aligned}
	\end{equation}
	The truncated Hilbert space is a tensor product between the qubit and qutrit spaces, i.e. $\mathcal{H} = \mathcal{H}_\textrm{qubit}\otimes\mathcal{H}_\textrm{qutrit}$, yielding a total dimension of six. The eigenstates are likewise tensor product states such that $\ket{ij} = \ket{i}_{1}\otimes\ket{j}_{2}$ denotes the combined state of the qubit/qutrit system\footnote{Subscript 1 denotes the qubit and subscript 2 denotes the qutrit.}. From now on, the identity operators in Eq. \eqref{eq:non_ket_representation_interaction_term} are omitted for simplicity. Finally, we write the interaction term step operators in the representation of the transmon eigenkets, from which one obtains
		\begin{equation}
		\begin{aligned}
		H = &\omega_{1}\ket{1}_{1}\bra{1}_{1} + \sum_{i=2}^{3}\omega_{i}\ket{i-1}_{2}\bra{i-1}_{2} + \omega_r b_{r}^{\dagger}b_{r}\\
		&+ g_{01}^{(1)}\Big(b_{r}\ket{1}_{1}\bra{0}_{1} + h.c.\Big)\\
		&+ g_{01}^{(2)}\Big(b_{r}\ket{1}_{2}\bra{0}_{2} + h.c.\Big)\\
		&+ g_{12}^{(2)}\Big(b_{r}\ket{2}_{2}\bra{1}_{2} + h.c.\Big). 	
		\label{eq:circuit_JC_Hamiltonian}
		\end{aligned}
		\end{equation}
		where $g_{m,m+1}^{(i)} = g_{i}\sqrt{m+1}$ denote nearest neighbour coupling constants for the qubit ($i = 1$) and the qutrit ($i = 2$).
	\subsection{Dispersive Jaynes-Cummings Hamiltonian}
	\label{sec:dispersive_JC_regime}
	In order to obtain a direct state-swapping interaction, one may utilize the dispersive Jaynes-Cummings regime in which the transmon frequencies are sufficiently detuned from the resonator such that $\frac{g}{\omega - \omega_{r}}\ll 1$ where the denominator is the relevant transmon energy-resonator detuning and $g$ is the corresponding nearest neighbour coupling constant. As a result, the interaction becomes a virtual exchange of photons through the resonator\cite{JerryChow}.\\
	The dispersive Hamiltonian is obtained by unitarily transforming the resonant Hamiltonian and expanding to second order using the Baker-Campbell-Hausdorff expansion such that \cite{2016arXiv160906796F}
	\begin{equation}
		\begin{aligned}
			H_{\textrm{dispersive}} = &\omega_{1}\ket{1}_{1}\bra{1}_{1} + \omega_{2}\ket{1}_{2}\bra{1}_{2} + \omega_{3}\ket{2}_{2}\bra{2}_{2} + \omega_r' b_{r}^{\dagger}b_{r}\\
			&+ \frac{g_{01}^{(1)}g_{01}^{(2)}(\Delta_{1} + \Delta_{2})}{2\Delta_{1}\Delta_{2}}\Big(\ket{01}\bra{10} + h.c.\Big)\\
			&+ \frac{g_{01}^{(1)}g_{12}^{(2)}(\Delta_{1} + \Delta_{3})}{2\Delta_{1}\Delta_{3}}\Big(\ket{11}\bra{02} + h.c.\Big)\\
			&+\frac{g_{01}^{(2)}g_{12}^{(2)}(\Delta_{2} - \Delta_{3})}{\Delta_{2}\Delta_{3}}\Big(b_{r}^{\dagger}b_{r}^{\dagger}\ket{0}_{2}\bra{2}_{2} + h.c.\Big)
			\label{eq:dispersive_Hamiltonian_all_interactions}
		\end{aligned}
	\end{equation}
	where $\Delta_{1} = \omega_{1} - \omega_{r}$, $\Delta_{2} = \omega_{2} - \omega_{r}$ and $\Delta_{3} = \omega_{3} - \omega_{2} - \omega_{r}$. The latter term in the Hamiltonian may be suppressed compared to the direct state-swapping term. Details of the suppression along with a presentation of the unitary transformation are presented in appendix \ref{sec:suppression}. The interaction term does no longer directly contain resonator degrees of freedom and hence the interaction serves to directly state-swap the transmons. The shifted resonator frequency, $\omega_{r}'$, is not discussed further since the resonator only acts as a mediator for the interaction and is omitted. Clearly, the coupling is strongly dependent on the relative detuning between the two transmons, being largest when the transmons are in resonance, i.e. $\Delta_{1} = \Delta_{3}$. With $\omega_1 + \omega_2 = \omega_3$, the first interaction $\ket{01}\rightarrow\ket{10}$ will be suppressed by conservation of energy. The final Hamiltonian of the main circuit (MC) in the dispersive regime reads
	\begin{equation}
		\begin{aligned}
			H_{\textrm{MC}} = &\omega_{1}\ket{1}_{1}\bra{1}_{1} + \omega_{2}\ket{1}_{2}\bra{1}_{2} + \omega_{3}\ket{2}_{2}\bra{2}_{2}\\
			&+ g\Big(\ket{11}\bra{02} + h.c.\Big)
			\label{eq:system_Hamiltonian}
		\end{aligned}
	\end{equation}
	with $g = \frac{g_{01}^{(1)}g_{12}^{(2)}}{\Delta}$ where $\Delta = \omega_1 - \omega_r$.\\
	We now review the validity of the applied approximations. One must take into account that the transmon-resonator detuning cannot exceed a value which disallows the rotating wave approximation of Eq. \eqref{eq:non_ket_representation_interaction_term}, but must also be large enough to justify the expansion of the Hamiltonian to second order in Eq. \eqref{eq:dispersive_Hamiltonian_all_interactions}. 
	In principle, we may get a more concrete handle on potential experimental parameters by further analysis of the concrete layout of the circuit design, and hence testing for which regimes our approximations will hold. However, noting that such an interaction has been experimentally realized \cite{2007Natur.449..443M}, we leave this for future studies. 
	Note that the interaction term in the Hamiltonian allows the qubit to excite the upper level of the qutrit, provided the qutrit is excited to its first level. Thus, we may control the population of the second excited state of the qutrit by modulating the qubit population. This modulation will be realized by adjusting a thermal bath temperature. Thus, the dynamics of Eq. \eqref{eq:system_Hamiltonian} are equivalent to the desired dynamics in \cite{PhysRevE.98.022118}. Now that we understand the dynamics of the main circuit, we will discuss the thermal baths.
	\section{Coupling of the main circuit to the thermal baths}
	\label{sec:thermal_transistor_Hamiltonian_and_open_quantum_system}
	The coupling of the main circuit to the thermal baths will be described perturbatively within the theory of open quantum systems with the nature of the interaction modelled as electromagnetic noise from the baths. To understand the coupling to the thermal baths, we now consider a conceptual model of the thermal baths in order to motivate the actual form of the Lindblad equation, specifically the collapse operators, for the thermal transistor.
	\subsection{Resistor model of the thermal baths}
	\label{sec:resistor_model_of_thermal_baths}
	Each thermal bath will be modelled as an LC circuit coupled to a noisy resistor using the formalism presented in \cite{quantum_noise}. The noisy resistor is modelled as a noiseless resistor coupled to a fluctuating voltage source as illustrated in Fig. \ref{fig:total_circuit_with_labels}. The system will generate a continous spectrum of electromagnetic noise due to the thermal agitation of Cooper pairs. The bath is described by a Hamiltonian of bosonic modes, $H_{\textrm{B}} = \sum_{k}\omega_{k}\Big(b_{k}^{\dagger}b_{k} + \frac{1}{2}\Big)$, whose population is dictated by temperature. Since the noise must pass the LC circuit (which acts as a band-pass filter), the voltage noise is restricted around the LC circuit resonant frequency, $\Omega$. Thus, only noise signals whose frequencies are in the vicinity of $\Omega$ will be transmitted through the LC circuit and participate in the main circuit interaction. From the formal treatment of the resistor model in \cite{quantum_noise}, one may derive the spectral density of the thermal bath circuit,
	\begin{equation}
		S(\omega) = \frac{1}{1 + Q^2(\frac{\omega}{\Omega} - \frac{\Omega}{\omega})^2}\frac{2R\omega}{1 - e^{-\omega/T}},
		\label{eq:spectral_density}
	\end{equation}
	where $T$ is the temperature of the bath, $Q$ is the LC circuit quality factor and $R$ is the resistance of the circuit resistor, a parameter fitted in experiments and effectively serves to scale the spectral densities \footnote{Using the conventional units presented in the circuit analysis of appendix \ref{sec:main_circuit_analysis}, capacitance and inductance have units of inverse energy. Hence, resistance is unitless and thus the unit of the spectral density becomes inverse time in natural units.}. The spectrum restricts the noise frequencies around $\Omega$ with its Lorentzian shape, depending on the Q-factor. This allows for suppressing unwanted transitions in the coupling with the main circuit.
	\subsection{Lindblad equation of the thermal transistor}
	\label{sec:lindblad_equation_of_the_thermal_transistor}
	In the following, we will introduce the main circuit-bath interaction terms phenomenologically. We assume that the system-bath coupling constant will be adequately small such that the main circuit eigenstates remain unperturbed, justifying the use of the Born approximation, ensuring the avoidance of correlations between the main circuit and the baths. We denote the capacitive coupling strength between the baths and the main circuit $\alpha$, and assume that it is equally strong for all baths.
	By tuning the bath band-pass filters, governed by Eq. \eqref{eq:spectral_density}, we restrict the interactions such that the source serves to excite/de-excite the lower state in the qutrit, the modulator to excite/de-excite the qubit and the drain to excite/de-excite the upper state in the qutrit to its ground state, see Fig. \ref{fig:conceptual_model}. Thus, the resonant frequencies of the thermal baths will be, respectively, $\Omega_{S} = \omega_2$, $\Omega_{M} = \omega_1$ and $\Omega_{D} = \omega_{3}$. This is illustrated in the interaction term,
	\begin{equation}
		\begin{aligned}
			H_{\textrm{MCB}} = &\sum_{k}\alpha\Big(b_{Sk}^{\dagger}\ket{0}_{2}\bra{1}_{2} + h.c.\Big)\\
			&+ \sum_{k}\alpha\Big(b_{Mk}^{\dagger}\ket{0}_{1}\bra{1}_{1} + h.c.\Big)\\
			&+ \sum_{k}\alpha\Big(b_{Dk}^{\dagger}\ket{0}_{2}\bra{2}_{2} + h.c.\Big)
			\label{eq:interaction_term_non_diagonal}
		\end{aligned}
	\end{equation}
	where $k$ is the bath summation index. $b_{\mu k}, b_{\mu k}^{\dagger}$ represent the bath creation/annihilation operators with $\mu\in\{S, M, D\}$, letting $S$, $M$ and $D$ (source, modulator and drain) refer to each thermal bath, according to Fig. \ref{fig:total_circuit_with_labels}. Note that in the weak coupling limit, $\alpha\ll g$. Since $T_S > T_D$, with $T_D$ being the drain temperature and $T_S$ being the source temperature, we expect the heat to only flow in the direction from the source to the drain. It is crucial that the upper level of the qutrit is only excited due to the interaction with the qubit, i.e. through the interaction in Eq. \eqref{eq:system_Hamiltonian}, since otherwise unwanted heat currents would pass. Hence, interactions such as $\sum_{k}\alpha\Big(b_{Dk}^{\dagger}\ket{0}_{2}\bra{1}_{2} + h.c.\Big)$ and $\sum_{k}\alpha\Big(b_{Sk}^{\dagger}\ket{0}_{2}\bra{2}_{2} + h.c.\Big)$ must be suppressed.
	Let $H_{\textrm{MC}} = \sum_{i=1}^{6}E_{i}\ket{E_{i}}\bra{E_{i}}$ denote the diagonal main circuit Hamiltonian with energies $E = \{\omega_{1} + \omega_{3}, \omega_{3} - g, \omega_{1}, \omega_{3} + g, \omega_{2}, 0\}$ and eigenstates $\ket{E_{i}}$. Following the procedure in \cite{PhysRevE.98.022118}, define the main circuit eigenoperators as
	\begin{equation}
	A_{\mu l}(\omega_{\mu l}) = \sum_{E_{i}-E_{j}=\omega_{\mu l}}\Pi(E_{j})D_{\mu}\Pi(E_{i})	
	\end{equation}
	for a fixed $\omega_{\mu l}$, with $l\in\{1,2,3\}$ denoting the three eigenoperators for each bath, $\mu$. $\Pi(E_{i})$ denotes a projection operator onto the eigenspace belonging to eigenvalue $E_{i}$. $D_{\mu}$ denotes main circuit operators in the interaction terms of Eq. \ref{eq:interaction_term_non_diagonal} and $\mu$ denotes the bath responsible for the transition. These eigenoperators serve to transition between energy eigenstates of the main circuit Hamiltonian. There exists nine eigenoperators for the main circuit which encapsulate the interactions of Eq. \eqref{eq:interaction_term_non_diagonal}, three for each bath, given in appendix \ref{sec:collapse_operators}. Noting that the $\Pi$ operators do not act in the Hilbert space of the baths, the diagonal representation reads\cite{PhysRevE.98.022118}
	\begin{equation}
	\begin{aligned}
	H_{\textrm{MCB}} = \sum_{\mu,l,k}\alpha\Big(b_{\mu k}^{\dagger}A_{\mu l}(\omega_{\mu l}) + h.c.\Big), &\quad \mu\in\{S, M, D\},\\
	&\quad l\in\{1,2,3\}.
	\end{aligned}
	\end{equation}
	It would be reasonable to use the eigenoperators of $H_{\textrm{MC}}$ as collapse operators in the Lindblad formalism due to their operation on energy eigenkets. We must therefore calculate the corresponding transition frequency for each collapse operator. This frequency will be the sum of all transition frequencies for all the transitions the collapse operator may be responsible for. Each transition rate between two states, $\ket{E_i}$ and $\ket{E_j}$, will be calculated using Fermi's golden rule, given by
	\begin{equation}
		\Gamma_{ij} = 2\pi\big\vert\bra{E_{j}}\alpha A_{\mu l}(\omega_{\mu l})\ket{E_{i}}\big\vert^{2}S(\omega_{\mu l})
	\end{equation}
	where $A_{\mu l}(\omega_{\mu l})$ is the collapse operator responsible for the transition between the two energy eigenkets. All transition rates are given in appendix \ref{sec:collapse_operators}.\\
	Using the collapse operators of appendix \ref{sec:collapse_operators}, the Lindblad equation reads, with $\rho \equiv \textrm{tr}_{\textrm{B}}(\rho_{\textrm{total}})$,
	\begin{widetext}
		\begin{equation}
			\begin{aligned}
				\frac{\partial\rho}{\partial t} = &\frac{1}{i}[H_{\textrm{MC}}, \rho]\\
				&+ \sum_{\mu l}\Big(\Gamma_{\mu l}(\omega_{\mu l})\Big[2A_{\mu l}(\omega_{\mu l})\rho A_{\mu l}^{\dagger}(\omega_{\mu l}) - \{A_{\mu l}^{\dagger}(\omega_{\mu l})A_{\mu l}(\omega_{\mu l}),\rho\}\Big]\\
				&+ \Gamma_{\mu l}(-\omega_{\mu l})\Big[2A_{\mu l}^{\dagger}(\omega_{\mu l})\rho A_{\mu l}(\omega_{\mu l}) - \{A_{\mu l}(\omega_{\mu l})A_{\mu l}^{\dagger}(\omega_{\mu l}),\rho\}\Big]\Big).
				\label{eq:system_Lindblad_equation}
			\end{aligned}
		\end{equation}
	\end{widetext}
	Since the spectral densities depend on temperature, it is possible to control the dynamics of the main circuit through temperature modulation of the thermal baths. Equipped with the Lindblad equation, we now move on to calculate the heat currents exchanged between the baths and the main circuit.	
	\section{Heat currents and transistor mechanisms}
	\label{sec:quantum_thermodynamics}
	The heat current will be defined as
	\begin{equation}
	J_{\mu} \equiv \langle L_{\mu}^{*}(H_{\textrm{MC}})\rangle.
	\label{eq:heat_current}
	\end{equation}
	as motivated from the quantum thermodynamical Heisenberg equation of motion.
	Using Eq. \eqref{eq:heat_current}, we may now proceed to calculate the heat currents exchanged between the source and qutrit, drain and qutrit, and modulator and qubit which all depend on the density matrix elements. Since the Lindblad equation is unidirectional, it takes any state to the invariant steady state, satisfying $\frac{\partial\rho}{\partial t} = 0$. Hence, we use the steady state density matrix elements to compute the heat currents, since these are invariants of the Lindblad equation of motion\cite{2018arXiv180306279N}.\\
	In the steady state, all off-diagonal elements of the density matrix vanish since the off-diagonal coherences will decay exponentially to zero\cite{PhysRevE.98.022118}. This motivates the introduction of the vectorization of the density matrix, analogously to what is done in \cite{PhysRevE.98.022118}, such that it is represented as $\ket{\rho} = [\rho_{11},...,\rho_{66}]^{T}$. The Lindblad equation may then be perceived as an operator equation where $\frac{\partial\ket{\rho}}{\partial t} = \mathcal{L}\ket{\rho}$ with $\mathcal{L}$ being the Lindblad superoperator. The objective is then to solve the eigenvalue equation $\mathcal{L}\ket{\rho} = 0$ by diagonalization of $\mathcal{L}$.
	The matrix representation of the eigenvalue problem is given by Eq. \eqref{eq:steady_state_matrix_equation}, with the first term commutator in Eq. \eqref{eq:system_Lindblad_equation} being zero in the steady state. In principle, the matrix in Eq. \eqref{eq:steady_state_matrix_equation} could be diagonalized analytically. However, with the spectral densities used in this article, the characteristic polynomial becomes considerably tedious and complicated and hence the process of diagonalization was done numerically. Specifically, the module NumPy in Python was used\cite{NumPy_diagonalization}.
	\subsection{Heat currents and amplification}
	\label{sec:heat_currents_and_amplification}
	By the heat current definition, Eq. \eqref{eq:heat_current}, the source current is, with $\alpha_{\mu l} \equiv \Gamma_{\mu l}(\omega_{\mu l})$ and $\beta_{\mu l} \equiv \Gamma_{\mu l}(-\omega_{\mu l})$,
	\begin{align}
		J_{S} =\langle L_{S}^{*}(H_{\textrm{MC}})\rangle &= (E_3 - E_2)(\alpha_{S1}\rho_{22} - \beta_{S1}\rho_{33})\nonumber\\
		&+ 2(E_6 - E_5)(\alpha_{S2}\rho_{55} - \beta_{S2}\rho_{66})\nonumber\\
		&+ (E_3 - E_4)(\alpha_{S3}\rho_{44} - \beta_{S3}\rho_{33}).
	\end{align}
	The heat currents of the modulator and drain are given in appendix \ref{sec:heat_currents_appendix}. As mentioned in the introduction of the article, amplification is a key ingredient to a transistor. The amplification factor is defined as
	\begin{equation}
		\alpha_{S,D} = \frac{\partial J_{S,D}}{\partial J_M}
		\label{eq:amplification}
	\end{equation}
	conforming to the convention in \cite{PhysRevE.98.022118, PhysRevB.88.094427,PhysRevA.97.052112}. For a given change in the modulating heat current, one would obtain amplification effects if $\alpha > 1$. It is crucial to note the dependence of the energy level differences in the qubit/qutrit on the amplification. Since the heat currents depend linearly on the energy differences of the qubit/qutrit, the degree of amplification depends naturally on the anharmonicity. This is numerically demonstrated in the following.
	\subsection{Numerical simulation}
	\label{sec:numerical_simulation}
\begin{figure}[h]
		\centering
		\includegraphics[width=1\columnwidth]{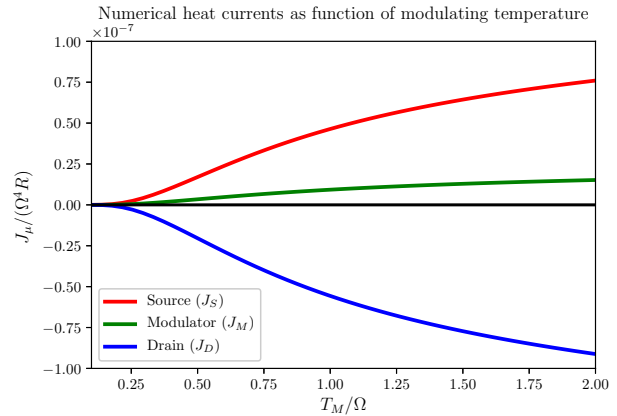}
		\caption{Numerical heat currents of the transistor circuit. When increasing the temperature of the modulator, qubit population rises and excites the qutrit, allowing for heat exchange between the source and drain. Heat currents and temperature are scaled according to the energy scale, $\Omega$, and the resistance of the baths, $R$, to obtain unitless entities. Simulation temperatures are $T_{D}/\Omega = 0.2$ and $T_{S}/\Omega = 2$. The anharmonicity is $\lambda = 4.0$ and in this high-anharmonicity regime, we observe the effect of amplification.}
		\label{fig:heat_currents}
	\end{figure}

\begin{figure}[h]
		\centering
		\includegraphics[width=1\columnwidth]{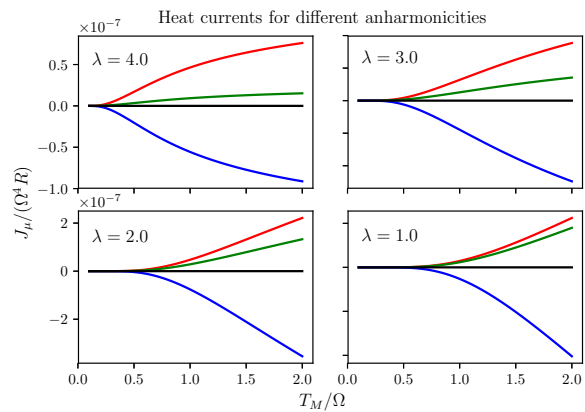}
		\caption{Heat currents as function of different values of anharmonicity. The rate of change of the modulator heat current increases as function of decreasing anharmonicity. In the limit of realistic values, i.e. $\lambda<1$, the amplification effect nearly vanishes.}
		\label{fig:amplification}
	\end{figure}

\begin{figure}[h]
		\centering
		\includegraphics[width=1\columnwidth]{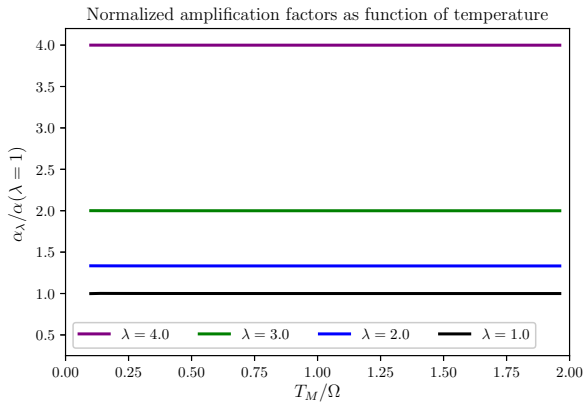}
		\caption{Normalized amplification factors as function of modulating temperature. Relative to the reference factor of $\lambda = 1.0$, the amplification increases by up to a factor of 4 with greater anharmonicities.}
		\label{fig:amp_factors}
	\end{figure}

	Typical first excitation energies of state-of-the-art transmons are on the order of $\omega_{01}/2\pi\sim 5$ GHz. A rough scale of transmon second excitation energy would be $\omega_{12}/2\pi\sim 4.8$ GHz with an anharmonicity of $200$ MHz\cite{PhysRevA.76.042319}. Therefore, we use relative parameters $\omega_1 = (\omega_3 - \omega_2), \omega_2 = 5.0\Omega$ and $\omega_3 = (10 - \lambda)\Omega$ where $\lambda$ is the anharmonicity. $\Omega$ defines the energy scale of the system (in the GHz regime for transmons). The internal coupling constant is chosen to be $g = 0.01\omega_1$ and a small bath coupling constant, $\alpha = 0.01g$, is chosen such that $\alpha\ll g$ to satisfy the weak-coupling limit of the Lindblad equation. All thermal baths are assumed to have quality factors $Q = 100$, which is a relatively high quality factor compared to what has been used in similar configurations \cite{2018NatPh..14..991R}. The source and drain baths have temperatures $T_{S} = 2\Omega$ and $T_{D} = 0.2\Omega$, respectively.
	Following the discussion in section \ref{sec:heat_currents_and_amplification}, we will demonstrate the effect of anharmonicity on the amplification. All constant simulation parameters are summarized in table \ref{table:simulation_parameters}.
	\begin{table}[h]
		\centering
		\begin{tabular}{@{}lll@{}}
			\toprule
			Parameter & Symbol & Numerical value  \\ \midrule
			Qubit/qutrit coupling constant & $g$ & $0.01\omega_{1}$ \\
			Thermal bath coupling constant & $\alpha$ & $0.01g$ \\ 
			Thermal bath Q-factor & $Q$ & $100$ \\
			Source temperature & $T_{S}$ & $2\Omega$ \\
			Drain temperature & $T_{D}$ & $0.2\Omega$ \\	 \bottomrule
		\end{tabular}
		\caption{Constant simulation parameters for the heat currents calculations.}
		\label{table:simulation_parameters}
	\end{table}
	Since $E_{i}\propto\Omega$ and $\alpha_{\mu l}\propto \Omega^3$, we divide the numerical heat currents by $\Omega^4$ obtain unitless quantities \footnote{With natural units, units of capacitance and inductance being inverse energy and units of spectral densities being inverse time, heat current has unit $[J_{\mu}] = \textrm{time}^{-4}$. Resistance is unitless in these conventions.}. The heat current is furthermore divided by the bath resistance, $R$, which is equal for all baths, serving only to scale the spectral density and in an experimental setting is a fitting parameter. First, we present the heat currents for a given anharmonicity, namely $\lambda= 4.0$. These are depicted in Fig. \ref{fig:heat_currents}. As can be seen, heat flows from the source into the drain reservoir for a given modulating heat current. Phyiscally, when $T_{M}/\Omega$ increases, the thermal population of photons in the modulator at the qubit energy is increased and hence the qubit population rises. As a result, the qutrit is excited at a higher rate and hence heat flows between the source and drain terminals of the transistor, since this flow requires the qutrit to occupy its highest energy level, restricted by the qubit population. The sum of all three heat currents equals 0, as it must by conservation of energy, which may be verified analytically and is also reflected in the numerical results. For an anharmonicity of $\lambda = 4.0$, the amplification effect seems evident from the simulation. The rates of change of source/drain heat currents are considerably larger as compared to the modulating heat current change for given temperature changes.
	Second, we present heat currents for four different anharmonicities as depicted in Fig. \ref{fig:amplification}, namely $\lambda = 1.0, 2.0, 3.0, 4.0$, to illustrate the effect of anharmonicity on amplification.
	As can be seen, decreasing the anharmonicity, i.e. toward more realistic transmon values, increases the modulating heat current. Physically, this effect may be understood as follows. The transferred heat current is blocked due to the constraint that the modulating qubit must excite the second level of the qutrit. The more identical the qutrit levels become, the more likely the modulating qubit will excite the levels of the qutrit and hence the heat current from the modulating bath becomes stronger. Using  Eq. \ref{eq:amplification}, we illustrate in Fig. \ref{fig:amp_factors} the amplification factors for each anharmonicity as function of the modulating temperature. The amplification factors are normalized relative to the amplification factor of the lowest anharmonicity and only the amplification factors of the source reservoir are calculated (one could equivalently calculate those for the drain reservoir).
	The amplification factor for the highest anharmonicity is observed to be around four times that of the more realistic anharmonicity. Thus, the effect of amplification is greatly reduced due to anharmonicity.
	\subsection{Switch mechanism}
	Other than the effect of amplification, we describe the switch mechanism of the transistor. As mentioned in the introduction of the article, a transistor must exhibit the ability to control the flow of signal between its terminals. In the case of the thermal transistor, the signal is the heat current between the source and drain terminals. In order to obtain a switch mechanism, two distinguishable on/off states must be defined. Thus, two modulating temperatures must be chosen for which the difference in heat currents is substantial enough to regard the transistor in an off state (vanishing heat current) at the lowest temperature and in an on state at the upper temperature. As an example, we consider the configuration in Fig. \ref{fig:heat_currents}. We define $T_{M}/\Omega = 0.25$ as an off state of the transistor, since the exchanged heat currents are practically vanishing, and $T_{M}/\Omega = 0.50$ as an on state. Thus, by modulating the temperature between these states, one  obtains a switch mechanism. From these considerations, the switch mechanism is evident: we have demonstrated that one may control the state of the transistor by adjusting the temperature of the modulator, realizing the ability to control heat current signals.
	\section{Conclusion, reflections and outlooks}
	Starting from the abstract thermal transistor model of \cite{PhysRevE.98.022118}, we provide a blueprint of an implementation in a superconducting circuit using circuit quantum electrodynamics. The main circuit, through which the heat flows from the source to the drain, was designed by utilizing the dispersive coupling regime of the Jaynes-Cummings model. By implementing the thermal baths as RLC circuits, the heat flow from the source to the drain is realized and shown to be adjustable by a third modulating bath. Effectively, this provided a switch mechanism between an {\it off}- and {\it on}-state of the transistor, depending on the temperature of the modulating terminal. 
	For experimental implementation, one would have to consider the detailed manufacturing and design of the resonator, transmon circuits and thermal baths in order to achieve the couplings and dynamics desired. 
	A potential way to realize the baths is to voltage-bias a pair of SINIS (superconductor-insulator-normal metal-insulator-superconductor) elements, which are NIS junctions in a series configuration. These elements would allow for temperature control of thermal baths and have been reported in a heat valve mechanism much like the switch mechanism presented here, although in a slightly different setting \cite{RevModPhys.78.217,2018NatPh..14..991R}.
	The amplification effect was demonstrated to decrease as function of decreasing anharmonicity, although not affecting the switch mechanism. Further improvements on this blueprint would be to consider the implementation of flux qubits which exhibit much higher anharmonicities\cite{2019ApPRv...6b1318K}.
	Theoretically, further improvements on the model could include a more thorough treatment of the thermal baths. Having a more detailed description of the spectral density and other physical properties of the thermal baths allows for the use of the Redfield master equation which in turn would yield a more in depth, microscopic understanding of the circuit dynamics. Furthermore, the effect of external electromagnetic field noise was not taken into account. In practice, the circuit will be subject to noise from its surroundings, giving rise to unwanted excitations and decays of the qubit or qutrit in the main circuit. External noise would obviously lead to unwanted heat currents to pass (or not to pass) and hence decreasing the efficiency of the transistor mechanism. However, state-of-the-art superconducting circuits now have extended lifetimes so that a proof-of-principle experiments should be possible with a near-term device. Another aspect is potential cross-talk between the thermal baths and related dissipative effects which would be present when considering the heat currents. These are topics for future investigations.

	\begin{acknowledgments}
		We wish to acknowledge Bao-qing Guo, Tong Liu, and Chang-shui Yu for their paper \cite{PhysRevE.98.022118} that was the original inspiration for our work. Furthermore, we acknowledge the insightful discussions with B. Karimi and J. Pekola which contributed to several aspects of the work. This work was supported by the Carlsberg Foundation.
	\end{acknowledgments}
	
	\appendix
	\section{Main circuit analysis and quantization} 
	\label{sec:main_circuit_analysis}
	
	
	With the flux nodes defined in section \ref{sec:main_circuit_Hamiltonian}, we ease notation by introducing matrix/vector notation such that $\vec{\phi} \equiv [\phi_1, \phi_2, \phi_3]^T$. The Lagrangian reads
	\begin{equation}
		\begin{aligned}
			L &= \frac{1}{2}\dot{\vec{\phi}}^{T}\underbrace{\begin{bmatrix}
					C_{1} + C_{g}^{(1)} & - C_{g}^{(1)} & 0\\
					- C_{g}^{(1)} & C_{2} + C_{g}^{(1)} + C_{g}^{(2)} & - C_{g}^{(2)} \\
					0 & - C_{g}^{(2)}  & C_{3} + C_{g}^{(2)}
			\end{bmatrix}}_{\textbf{C}}\dot{\vec{\phi}}\\
			&- \frac{1}{2L_{2}}\phi_{2}^{2}
			+ E_{J1}\cos\Big(\frac{2\pi}{\Phi_0}\phi_1\Big) + E_{J3}\cos\Big(\frac{2\pi}{\Phi_0}\phi_3\Big)
		\end{aligned}
	\end{equation}
	where $\mathbf{C}$ is the capacitance matrix. The above circuit parameters are depicted in Fig. \ref{fig:total_circuit_with_labels}. By writing out the capacitive terms explicitly, one may verify the equivalence of what would be obtained through the usual application of the Kirchhoff circuit rules. A quick procedure on how to write down the capacitance matrix easily is given in reference \cite{StigRasmussen}.
	Using a Legendre transformation, define the conjugate momentum as $\vec{p} = \frac{\partial L}{\partial \dot{\vec{\phi}}}$ and, by noting the invertibility of the capacitance matrix\cite{StigRasmussen}, the Hamiltonian reads
	\begin{equation}
		\begin{aligned}
			H &= \frac{1}{2}\vec{p}^{T}\textbf{C}^{-1}\vec{p} + \frac{1}{2L_{2}}\phi_{2}^{2} - E_{J1}\cos\Big(\frac{2\pi}{\Phi_0}\phi_1\Big)\\
			&- E_{J3}\cos\Big(\frac{2\pi}{\Phi_0}\phi_3\Big).
			\label{Hamiltonian_non_conventional_units}
		\end{aligned}
	\end{equation}
	The kinetic term reads
	\begin{equation}
		T = \frac{1}{2}\sum_{i=1}^{3}C_{ii}^{-1}p_{i}^2 + C_{12}^{-1}p_{1}p_{2} + C_{23}^{-1}p_{2}p_{3} + \underbrace{C_{13}^{-1}p_{1}p_{3}}_{\approx\text{0}}
		\label{eq:kinetic_term}
	\end{equation}
	where $C_{ij}^{-1}$ denotes the inverse capacitance matrix elements given by Eq. \ref{eq:inverse_capacitance_matrix}. The spatial separation of the transmon circuits in such a resonator architecture allows the direct capacitive cross-coupling between the transmons to be neglected \cite{PhysRevA.75.032329}. In principle, the cross-coupling term may be included, as studied in \cite{PhysRevB.73.125336}. In the architecture studied here, however, the cross-coupling is assumed to be negligible. The circuit Hamiltonian is conveniently expressed in terms of effective quantities using the canonical transformations $C^{-1}\rightarrow\frac{8}{(2e)^2}C^{-1}$, $E_{Ci} \equiv C_{ii}^{-1}$ and $E_{Li} \equiv \frac{1}{2L_{i}}$, with $e$ being the elementary charge. Capacitive terms then reduce to $4E_{Ci}\Big(\frac{p_{i}}{2e}\Big)^2$ where $E_{Ci}$ will be the effective energy at the $i$'th node in the circuit and $p_{i}/2e$ will denote the number of Cooper pairs stored at the corresponding flux node\cite{StigRasmussen}. Finally, the flux and its conjugate momentum are converted to dimensionless entities by letting $\frac{2\pi}{\Phi_0} = 1$ which leads to $2e = 1$, letting the charge of the Cooper pairs equal unity. Capacitance and inductance then obtain units of inverse energy, yielding a Hamiltonian in units of energy. Using these conventions, Eq. \ref{Hamiltonian_non_conventional_units} reads
	\begin{equation}
		\begin{aligned}
			H = &4E_{C1}p_{1}^2 - E_{J1}\cos(\phi_1) + 4E_{C3}p_{3}^2 - E_{J3}\cos(\phi_3)\\
			&+ 4E_{C2}p_{2}^2 + E_{L2}\phi_{2}^2 + 8C_{12}^{-1}p_{1}p_{2} + 8C_{23}^{-1}p_{2}p_{3}.
			\label{eq:non_transmon_Hamiltonian}
		\end{aligned}
	\end{equation}
	When operated in the transmon regime with $E_C/E_J\ll 1$, a perturbative expansion of the cosine terms in Eq. \ref{eq:non_transmon_Hamiltonian} yields \cite{PhysRevA.76.042319}
	\begin{align}
		H \approx &\overbrace{4E_{C1}p_{1}^2 + \frac{E_{J1}}{2}\phi_{1}^2 - \frac{E_{J1}}{24}\phi_{1}^4}^{\textrm{Transmon 1}}\nonumber\\
		&+ \overbrace{4E_{C3}p_{3}^2 + \frac{E_{J3}}{2}\phi_{3}^2 - \frac{E_{J3}}{24}\phi_{3}^4}^{\textrm{Transmon 2}} + \overbrace{4E_{C2}p_{2}^2 + E_{L2}\phi_{2}^2}^{\textrm{Resonator}}\nonumber\\
		&+ \overbrace{8C_{12}^{-1}p_{1}p_{2} + 8C_{23}^{-1}p_{2}p_{3}}^{\textrm{Couplings}}	
		\label{eq:main_circuit_classical_Hamiltonian}
	\end{align}
	neglecting constant terms of the Josephson tunnelling energy, $E_{Ji}, i\in\{1,3\}$. The first two collected terms are perceived as two harmonic oscillators with an anharmonic perturbation, the third collected term as a harmonic oscillator and the last collected terms as mutual couplings between the transmons and the resonator.\\
	The main circuit is canonically quantized by introducing the operators $p_i\rightarrow\hat{p}_i$ and $\phi_i\rightarrow\hat{\phi}_i$ with canonical commutation relation $[\hat{\phi_i},\hat{p_j}] = i\delta_{ij}$\footnote{Note the analogy with the position and momentum commutation relation, $[\hat{x},\hat{p}] = i$.}. Furthermore, we introduce step operators defined by \cite{KasperSangild}
	\begin{align}
		&\hat{a}_{i} = \frac{1}{2}\Big(\frac{E_{Li} + \frac{E_{Ji}}{2}}{E_{Ci}}\Big)^{1/4}\hat{\phi_i} + i\Big(\frac{E_{Ci}}{E_{Li} + \frac{E_{Ji}}{2}}\Big)^{1/4}\hat{p_i}\\
		&\hat{a}_{i}^{\dagger} = \frac{1}{2}\Big(\frac{E_{Li} + \frac{E_{Ji}}{2}}{E_{Ci}}\Big)^{1/4}\hat{\phi_i} - i\Big(\frac{E_{Ci}}{E_{Li} + \frac{E_{Ji}}{2}}\Big)^{1/4}\hat{p_i}.
	\end{align}
	From now on, we assume the operator nature of $a$ and $b$ are implied such that we lose the hat-notation. After some algebra, the Hamiltonian, Eq. \ref{eq:main_circuit_classical_Hamiltonian}, reduces to Eq. \ref{eq:non_truncated_circuit_Hamiltonian}.
	\section{Inverse capacitance matrix elements}
	\label{section:circuit_analysis_matrices}
	The inverse matrix elements are given by
	\begin{equation} 
	\begin{aligned}
		C_{11}^{-1} &= (C_2C_3 + C_2C_{g}^{(2)} + C_3C_{g}^{(1)} + C_3C_{g}^{(2)} + C_4C_{g}^{(2)})/\lambda\\
		C_{12}^{-1} &= C_{g}^{(1)}(C_3 + C_{g}^{(2)})/\lambda\\
		C_{13}^{-1} &= C_{g}^{(1)}C_{g}^{(2)}/\lambda\\
		C_{22}^{-1} &= (C_1 + C_{g}^{(1)})(C_3 + C_{g}^{(2)})/\lambda\\
		C_{23}^{-1} &= C_{g}^{(2)}(C_1 + C_{g}^{(1)})/\lambda\\
		C_{33}^{-1} &= (C_1C_2 + C_1C_{g}^{(1)} + C_1C_{g}^{(2)} + C_2C_{g}^{(1)} + C_{g}^{(1)}C_{g}^{(2)})/\lambda
		\label{eq:inverse_capacitance_matrix}
	\end{aligned}
	\end{equation}
	with $\lambda = C_1(C_2C_3 + C_2C_{g}^{(2)} + C_3C_{g}^{(1)} + C_3C_{g}^{(2)} + C_{g}^{(1)}C_{g}^{(2)}) + C_2(C_3C_{g}^{(1)} + C_{g}^{(1)}C_{g}^{(2)}) + C_3C_{g}^{(1)}C_{g}^{(2)}$.	Since the inverse capacitance matrix is symmetric, $C_{ij}^{-1} = C_{ji}^{-1}$ and hence the above equations encapsulate all matrix elements.
	\section{Collapse operators of the Lindblad equation}
	\label{sec:collapse_operators}
	The collapse operators of the Lindblad equation read\cite{PhysRevE.98.022118}
	\begingroup
	\allowdisplaybreaks
\begin{equation}
\begin{aligned}
		A_{S1} &= \frac{1}{\sqrt{2}}\ket{E_{3}}\bra{E_{2}}, \quad \omega_{S1} = E_2 - g\\
		A_{S2} &= \ket{E_{6}}\bra{E_{5}}, \quad \omega_{S2} = E_2\\
		A_{S3} &= \frac{1}{\sqrt{2}}\ket{E_{3}}\bra{E_{4}}, \quad \omega_{S3} = E_2 + g\\
		A_{M1} &= \ket{E_{6}}\bra{E_{3}}, \quad \omega_{M1} = E_1\\
		A_{M2} &= \frac{1}{\sqrt{2}}\Big(\ket{E_{5}}\bra{E_{2}} + \ket{E_{4}}\bra{E_{1}}\Big), \quad \omega_{M2} = E_1 - g\\
		A_{M3} &= \frac{1}{\sqrt{2}}\Big(\ket{E_{5}}\bra{E_{4}} - \ket{E_{2}}\bra{E_{1}}\Big), \quad \omega_{M3} = E_1 + g\\
		A_{D1} &= \frac{1}{\sqrt{2}}\ket{E_{6}}\bra{E_{2}}, \quad \omega_{D1} = E_3 - g\\
		A_{D2} &= -\frac{1}{\sqrt{2}}\ket{E_{6}}\bra{E_{4}}, \quad \omega_{D2} = E_3 + g\\
		A_{D3} &= \ket{E_{3}}\bra{E_{1}}, \quad \omega_{D3} = E_3.
\end{aligned}
\end{equation}		
	\endgroup
	The transition rates of the Lindblad collapse operators are calculated using Fermi's golden rule and are given by
	\begingroup
	\allowdisplaybreaks
	\begin{equation}
	\begin{aligned}
		\Gamma_{S1} &= \Gamma_{23} = \pi\alpha^{2}S(\omega_{S1}),\quad
		\Gamma_{S2} = \Gamma_{56} = 2\pi\alpha^{2}S(\omega_{S2})\\
		\Gamma_{S3} &= \Gamma_{43} = \pi\alpha^{2}S(\omega_{S3}),\quad
		\Gamma_{M1} = \Gamma_{36} = 2\pi\alpha^{2}S(\omega_{M1})\\
		\Gamma_{M2} &= \Gamma_{25} + \Gamma_{14} = 2\pi\alpha^{2}S(\omega_{M2})\\
		\Gamma_{M3} &= \Gamma_{45} + \Gamma_{12} = 2\pi\alpha^{2}S(\omega_{M3})\\
		\Gamma_{D1} &= \Gamma_{26} = \pi\alpha^{2}S(\omega_{D1}),\quad
		\Gamma_{D2} = \Gamma_{46} = \pi\alpha^{2}S(\omega_{D2})\\
		\Gamma_{D3} &= \Gamma_{13} = 2\pi\alpha^{2}S(\omega_{D3}).
	\end{aligned}
	\end{equation}
	\endgroup
	\section{Suppression of interaction term in BCH expansion of the main circuit Hamiltonian}
	\label{sec:suppression}
	The unitary transformation used to eliminate resonant terms of Eq. \ref{eq:circuit_JC_Hamiltonian} reads
	\begin{equation}
	\begin{aligned}
	U = \exp\Big(&\lambda_{0}^{(1)}(b_{r}^{\dagger}\ket{0}_{1}\bra{1}_{1} - h.c.) + \lambda_{0}^{(2)}(b_{r}^{\dagger}\ket{0}_{2}\bra{1}_{2} - h.c.)\\
	&+ \lambda_{1}^{(2)}(b_{r}^{\dagger}\ket{1}_{2}\bra{2}_{2} - h.c.)\Big).
	\end{aligned}
	\end{equation}
	where $\lambda_{i}^{(k)} = \frac{g_{i,i+1}^{(k)}}{\omega_{i,i+1}^{(k)}-\omega_{r}}\ll 1$ in the dispersive regime. Expansion to second order in $\lambda_{i}^{(k)}$ using the Baker-Campbell-Hausdorff expansion yields Eq. \ref{eq:dispersive_Hamiltonian_all_interactions}. Since the latter term is unwanted in the transistor mechanism, the inequality
	\begin{equation}
	\begin{aligned}
	\frac{g_{01}^{(1)}g_{12}^{(2)}(\Delta_{1} + \Delta_{3})}{2\Delta_{1}\Delta_{3}}\gg \frac{g_{01}^{(2)}g_{12}^{(2)}(\Delta_{2} - \Delta_{3})}{\Delta_{2}\Delta_{3}}
	\end{aligned}
	\end{equation}
	must be satisfied. The detuning of the resonator frequency relative to the first excitation energy of the qutrit may be expressed as $\omega_{2} = \beta\omega_{r}$ where $\beta$ parametrizes the detuning. Using the resonant conditions where $\Delta_{1} = \Delta_{3}$ and the qubit/qutrit energy parametrizations of section \ref{sec:numerical_simulation}, namely $\omega_{3} = (10-\lambda)\Omega$ and $\omega_{2} = 5\Omega$, one obtains the inequality
	\begin{equation}
	\frac{g_{1}}{g_{3}}\gg \frac{\lambda/5}{1-\beta}.
	\end{equation}
	Thus, for a given detuning parameter $\beta$, the suppression of the unwanted transition depends on the anharmonicity. For a given detuning, the coupling
	\begin{equation}
	\frac{g_{01}^{(1)}g_{12}^{(2)}(\Delta_{1} + \Delta_{3})}{2\Delta_{1}\Delta_{3}}\Big(\ket{11}\bra{02} + h.c.\Big)
	\end{equation}
	is invariant under relative changes in $g_{1}$ and $g_{2}$ since $g_{01}^{(1)}\propto g_{1}$ and $g_{12}^{(2)}\propto g_{3}$. As an example, consider $\omega_{r}/2\pi = 3.5$ GHz and $\omega_{2}/2\pi = 5.0$ GHz which yields $\beta = 5/7$. One obtains
	\begin{equation}
	\frac{g_{1}}{g_{3}}\gg \frac{7}{10}\lambda.
	\label{eq:inequality}
	\end{equation}
	Thus, by utilizing the flexible scaling of the coupling capacitances, one may adjust $g_{1}$ and $g_{3}$ to satisfy Eq. \ref{eq:inequality} which suppresses the unwanted transition.
	\section{Steady state Lindblad matrix equation}
	\label{sec:steady_state_Lindblad_equation}
	The steady state Lindblad equation reads $\mathcal{L}\ket{\rho} = 0$ and the corresponding matrix representation of this equation reads
	\begin{equation}
		\begin{bmatrix}
			\gamma_1 & \beta_{M3} & 2\beta_{D3} & \beta_{M2} & 0 & 0 \\
			\alpha_{M3} & \gamma_2 & \beta_{S1} & 0 & \beta_{M2} & \beta_{D1} \\
			2\alpha_{D3} & \alpha_{S1} & \gamma_3 & \alpha_{S3} & 0 & 2\beta_{M1} \\
			\alpha_{M2} & 0 & \beta_{S3} & \gamma_4 & \beta_{M3} & \beta_{D2} \\
			0 & \alpha_{M2} & 0 & \alpha_{M3} & \gamma_5 & 2\beta_{S2} \\
			0 & \alpha_{D1} & 2\alpha_{M1} & \alpha_{D2} & 2\alpha_{S2} & \gamma_6
		\end{bmatrix}
		\begin{bmatrix}
			\rho_{11} \\
			\rho_{22} \\
			\rho_{33} \\
			\rho_{44} \\
			\rho_{55} \\
			\rho_{66} \\
		\end{bmatrix}
		= 0
		\label{eq:steady_state_matrix_equation}
	\end{equation}
	with
	\begingroup
	\allowdisplaybreaks
	\begin{align}
		\gamma_1 &= - \alpha_{M2} - \alpha_{M3} - 2\alpha_{D3}\nonumber\\
		\gamma_2 &= - \alpha_{S1} - \alpha_{M2} - \beta_{M3} - \alpha_{D1}\nonumber\\
		\gamma_3 &= - \beta_{S1} - \beta_{S3} - 2\alpha_{M1} - 2\beta_{D3}\nonumber\\
		\gamma_4 &= - \alpha_{S3} - \beta_{M2} - \alpha_{M3} - \alpha_{D2}\nonumber\\
		\gamma_5 &= - 2\alpha_{S2} - \beta_{M2} - \beta_{M3}\nonumber\\
		\gamma_6 &= - 2\beta_{S2} - 2\beta_{M1} - \beta_{D1} - \beta_{D2}\nonumber.
	\end{align}
	\endgroup
	and $\alpha_{\mu l} = \Gamma_{\mu l}(\omega_{\mu l})$ and $\beta_{\mu l} = \Gamma_{\mu l}(-\omega_{\mu l})$ for $\mu\in\{S, M, D\}$ and $l\in\{1,2,3\}$.
	\section{Heat currents of modulator and drain baths}
	\label{sec:heat_currents_appendix}
	Using Eq. \ref{eq:heat_current}, we calculate the heat currents of the modulator and drain, analogously to what was done in the main text for the source.
	We obtain
	\begingroup
	\allowdisplaybreaks
	\begin{equation}
		\begin{aligned}
			J_M = &2(E_6 - E_3)(\alpha_{M1}\rho_{33} - \beta_{M1}\rho_{66})\\
			&+ (E_4 - E_1)(\alpha_{M2}\rho_{44} - \beta_{M2}\rho_{11})\\
			&+ (E_5 - E_2)(\alpha_{M2}\rho_{22} - \beta_{M2}\rho_{55})\\
			&+ (E_2 - E_1)(\alpha_{M3}\rho_{11} - \beta_{M3}\rho_{22})\\
			&+ (E_5 - E_4)(\alpha_{M3}\rho_{44} - \beta_{M3}\rho_{55})
			\label{eq:middle_bath_heat_currents}
		\end{aligned}
	\end{equation}
	\endgroup
	and
	\begin{equation}
		\begin{aligned}
			J_D = &(E_6 - E_2)(\alpha_{D1}\rho_{22} - \beta_{D1}\rho_{66})\\
			&+ (E_6 - E_4)(\alpha_{D2}\rho_{44} - \beta_{D2}\rho_{66})\\
			&+ 2(E_3 - E_1)(\alpha_{D3}\rho_{11} - \beta_{D3}\rho_{33}).
			\label{eq:right_bath_heat_currents}
		\end{aligned}
	\end{equation}
	
\bibliography{ms}
\end{document}